\documentclass[aps,prl,reprint,superscriptaddress]{revtex4-1}

\usepackage[]{changes}

\usepackage{url}
\def\UrlBreaks{\do\/\do-}

\usepackage{slashed}
\usepackage{amssymb}
\usepackage{graphicx}
\usepackage{amsmath}
\usepackage{epsfig}
\usepackage{multirow}
\usepackage{color}
\usepackage[pdftitle={Evaluation of multi-loop multi-scale Feynman integrals for precision physics},colorlinks=true, allcolors=blue,breaklinks,pdfdisplaydoctitle]{hyperref}
\makeatletter
\g@addto@macro{\UrlBreaks}{\UrlOrds}
\g@addto@macro{\UrlBreaks}{%
\do\/\do\d%
}
\makeatother
\newcommand{\MZ}{M_{\rm Z}}

\newcommand{\mz}{m_{\rm Z}}
\newcommand{\MW}{M_{\rm W}}

\newcommand{\mw}{m_{\rm W}}

\newcommand{\mt}{M_{\rm t}}
\newcommand{\lmt}{m_{\rm t}}

\newcommand{\as}{\alpha_{\rm s}}

\begin{document}

\title{
Evaluation of multi-loop multi-scale Feynman integrals for precision physics
}
\author{Ievgen Dubovyk}
\affiliation{Institute of Physics, University of Silesia, Katowice, Poland}
\author{ Ayres Freitas}
\affiliation{Pittsburgh Particle physics, Astrophysics \& Cosmology Center
(PITT PACC),\\ Department of Physics \& Astronomy, University of Pittsburgh, Pittsburgh, PA 15260, USA}
\author{Janusz Gluza}
\affiliation{Institute of Physics, University of Silesia, Katowice, Poland} 
\author{Krzysztof Grzanka}
\affiliation{Institute of Physics, University of Silesia, Katowice, Poland} 
\author{Martijn Hidding}
\affiliation{Department of Physics and Astronomy, Uppsala University, SE-75120 Uppsala, Sweden} 
\author{Johann Usovitsch}
\affiliation{Theoretical Physics Department, CERN, 1211 Geneva, Switzerland}  

\date{\today}
{\footnotesize
\hspace*{0pt}\hfill \\
\hspace*{0pt}\hfill\parbox{100pt}
{CERN-TH-2021-230\\
 UUITP-66/21
}
}

\begin{abstract} 
Modern particle physics is increasingly becoming a precision science that relies on advanced theoretical predictions for the analysis and interpretation of experimental results. The planned physics program at the LHC and future colliders will require three-loop electroweak and mixed electroweak-QCD corrections to single-particle production and decay processes and two-loop electroweak corrections to pair production processes. This article presents a new semi-numerical approach to multi-loop multi-scale Feynman integrals calculations which will be able to fill the gap between rigid experimental demands and theory. The approach is based on differential equations with boundary terms specified at Euclidean kinematic points. These Euclidean boundary terms can be computed numerically with high accuracy using sector decomposition or other numerical methods. They are then mapped to the physical kinematic configuration with a series solution of the differential equation system. The method is able to deliver 8 or more digits precision, and it has a built-in mechanism for checking the accuracy of the obtained results. Its efficacy is illustrated with state-of-the-art examples for three-loop self-energy and vertex integrals and two-loop box integrals.

\medskip
(Dedicated to the memory of Tord Riemann)
\end{abstract}

 
\maketitle

\paragraph{{\rm I.} Introduction.}

With the discovery of the Higgs boson at the Large Hadron Collider (LHC), all building blocks of the Standard Model (SM) have been experimentally confirmed, with the only exception of the Higgs self-coupling, which still awaits direct measurement. However, the SM does not account for important phenomena such as dark matter and the matter-antimatter asymmetry, so that physics beyond the SM is needed. It is reasonable to expect that this new physics couples to the electroweak and/or Higgs sector of the SM, since there are important model-building constraints for couplings to the strong force \footnote{A new physics particle with renormalizable QCD couplings but no weak or Higgs interactions would be stable and thus cosmologically excluded.}. 
  
Therefore, possible evidence for such new physics can be explored in precision studies of electroweak and Higgs physics at the high-luminosity run of the LHC (HL-LHC) or one of several proposed future $e^+e^-$ colliders: FCC-ee \cite{Abada:2019zxq}, CEPC~\cite{CEPCStudyGroup:2018ghi}, ILC~\cite{Baer:2013cma,Bambade:2019fyw}, CLIC~\cite{Linssen:2012hp,Charles:2018vfv}. 
Through their high integrated luminosities of several ab$^{-1}$, these machines will be sensitive to very small deviations between the measured value and the SM expectation for a given observable. Thus they can probe extremely feebly coupled new particles or very large new physics scales of tens of TeV.

The SM predictions for these precision analyses are obtained by computing higher-order quantum corrections. At the HL-LHC, some of the most interesting precision studies are Higgs boson production and lepton pair (Drell-Yan) production. For the former, one of the largest sources of theoretical uncertainty 
stems from
mixed QCD-electroweak corrections \cite{Anastasiou:2008tj,Anastasiou:2016cez}. While some partial results at this order have been computed \cite{Bonetti:2017ovy,Anastasiou:2018adr,Bonetti:2020hqh,Becchetti:2020wof,Becchetti:2021axs}, 
contributions from electroweak diagrams with internal top quarks, both for 3-loop Higgs production and 2-loop Higgs+jet production, are still needed to complete this missing piece.
For Drell-Yan production, 2-loop electroweak corrections for the full process $pp \to \ell^+\ell^-$, not just on the Z-boson resonance, are important since LHC measurements cover a broad range of invariant mass \cite{ATLAS:2018gqq,CMS:2018ktx}.

Similarly, electroweak 2-loop corrections for several different pair production processes will be essential for the physics goals of future $e^+e^-$ colliders \cite{Freitas:2019bre}: $e^+e^- \to W^+W^-$, $e^+e^- \to ZH$, and $e^+e^- \to f\bar{f}$. Measurements of these cross-sections will allow us to determine the W-boson mass with high precision, constrain anomalous couplings between gauge bosons and/or the Higgs boson, and probe heavy neutral vector bosons ($Z'$ bosons). Currently, some results for mixed QCD-electroweak 2-loop corrections are available \cite{Gong:2016jys,Sun:2016bel,Heller:2020owb,Bonciani:2021zzf}, but so far no complete electroweak 2-loop calculation for any pair production process has been carried out. Even higher-order corrections will be needed for studies of Z-boson production and decay at these future $e^+e^-$ colliders, as well as the indirect prediction of the $W$ boson mass from the Fermi constant. To match the expected experimental precision, 3-loop and partial 4-loop self-energy and vertex corrections will be required \cite{Blondel:2018mad,Freitas:2019bre}, which is one order of perturbation theory beyond the current state of the art \cite{Dubovyk:2019szj}. 

It should be emphasized that these are loop corrections in the full SM, involving many massive particles inside the loops. The currently most advanced techniques for analytically computing such multi-loop Feynman integrals first reduce them to a small set of master integrals, which then are solved by constructing suitable differential equations (DEs), see Ref.~\cite{Kotikov:2021tai} for a recent review. Both of these steps require integration-by-parts (IBP) equation systems \cite{Chetyrkin:1981qh,Laporta:2000dsw} that become intractable for multi-loop integrals with many masses. Instead, one must resort to numerical integration techniques.

The recent calculation of full 2-loop corrections to Z-boson production and decay \cite{Dubovyk:2016aqv,Dubovyk:2018rlg,Dubovyk:2019szj} made use of numerical evaluations based on sector decomposition (SD) \cite{Binoth:2000ps,Borowka:2017idc,Borowka:2018goh,Smirnov:2008py,Smirnov:2015mct}  and Mellin-Barnes (MB) representations \cite{Gluza:2007rt,Czakon:2005rk, Usovitsch-phd:2018qmt,Czakon:2005rk,Dubovyk:2016ocz,Usovitsch:2018shx,Dubovyk:2016aqv}. However, these methods require large amounts of computing resources and do not always converge to the required level of accuracy, so that a straightforward extension to more loops and/or legs is not possible. Due to numerical cancellations between individual loop integrals, at least 8 digits precision are required in many cases for practical applications \cite{Dubovyk:2018rlg}.
Another interesting approach, which allows to tackle a wider class of problems, evaluates a set of master integrals by numerically solving a DE system, either in terms of kinematic parameters \cite{Mandal:2018cdj,Czakon:2020vql} or in terms of an auxiliary flow variable \cite{Liu:2017jxz,Liu:2018dmc,Bronnum-Hansen:2020mzk}.

This article introduces an efficient but still very general
approach that can be applied to many challenging 2- and 3-loop problems with multiple mass and momentum scales \footnote{In principle, it is also applicable at higher loop orders, but we have not studied any examples beyond three loops.}.
The key elements are a system of DEs, with boundary terms evaluated at one or more Euclidean (space-like) kinematic points (which can be reliably determined to high precision with numerical methods). The DEs are then solved, using series expansions, to obtain the final result at the physical Minkowski (time-like) kinematic point. This approach,which is already fully automated in its main parts, will be described in more detail in the next section.
In section III we will apply this technique to examples of SM self-energy and vertex Feynman integrals that occur in three-loop Z-decay corrections.
The chosen examples are very difficult to evaluate with other analytical or numerical methods.
A summary and outlook are given in the final section. The supplemental material included with this submission contains input parameters for the example integrals, additional examples for the application of our calculation technique, e.g. a 2-loop four-scale box diagram, and miscellaneous implementational details and remarks for a more detailed discussion of the method.

\paragraph{{\rm II.} Description of the method.}
Solving Feynman integrals from DEs is an approach initiated in the last decade of the last century \cite{Kotikov:1990kg,Kotikov:1991hm,Kotikov:1991pm,Remiddi:1997ny}. Many families of Feynman integrals admit a choice of master integrals for which the system of DEs has a particularly simple ``canonical'' form \cite{Henn:2013pwa}, which in many cases can be straightforwardly solved in terms of multiple polylogarithms.

More generally, not all Feynman integrals are of polylogarithmic type, and it can become increasingly difficult to find a closed set of analytic functions in terms of which the DEs can be solved. In such cases, it is often still possible to find precise numerical solutions.  For example, the DE system can be integrated numerically \cite{Mandal:2018cdj,Czakon:2020vql,Liu:2017jxz,Liu:2018dmc,Bronnum-Hansen:2020mzk}.
In this work, we use the approach of iterated series expansions, which is both fully automatable and numerically efficient.

Let us give a brief overview of the method. Consider a basis of master integrals (MIs), $\vec F(x,\epsilon)$, depending on a single scale $x$. We work in dimensional regularization, with $D=4-2\epsilon$ space-time dimensions. We may then derive DEs of the form:
\begin{equation}
    \frac{d}{d x} \vec F(x,\epsilon) = \hat M(x,\epsilon) \vec F(x,\epsilon)\,,
\end{equation}
where $\hat M(x,\epsilon)$ is a block-triangular matrix. Each block is associated with a sector of integrals. If we denote such a sector by $\vec{f}_i(x,\epsilon)$, we can decompose the DEs in the form:
\begin{equation} \label{eq:general-de}
    \frac{d}{d x} \vec{f}_i(x,\epsilon) = M_i(x,\epsilon) \vec{f}_i(x,\epsilon) + B_i(x,\epsilon) \vec{g}_i(x,\epsilon)\,,
\end{equation}
where $M_i(x,\epsilon)$ denotes the diagonal block of $\hat M(x,\epsilon)$ corresponding to the sector $i$, and $B_i(x,\epsilon) \vec{g}_i(x,\epsilon)$ captures the off-diagonal terms. One can then expand the integrals and matrices in $\epsilon$:
\begin{align}
    \vec f_i(x,\epsilon) &= \sum_{j=-k}^\infty \vec f_i^{(j)}(x,\epsilon)\,  \epsilon^j\,, \notag \\
    M_i(x,\epsilon) &= \sum_{j=0}^\infty M_i^{(j)}(x,\epsilon)\, \epsilon^j\,,
\end{align}
and solve the system order by order in $\epsilon$. For a given basis, the condition that $M_i(x,\epsilon)$ is finite in $\epsilon$ is not always manifest. We obtain such a form by rescaling individual master integrals with powers of $\epsilon$, until the matrix is finite.
For further ideas and software to help facilitate the choice of MIs, see Refs.~\cite{Prausa:2017ltv,Gituliar:2017vzm,vonManteuffel:2017hms,Adams:2017tga,Henn:2020lye,Dlapa:2020cwj,Usovitsch:2020jrk,Smirnov:2020quc}.
 
The DE system in eq.~(\ref{eq:general-de}) fixes the master integrals up to some boundary conditions.
It turns out that, in the case of our automated DE approach, a convenient choice for the boundary terms are MIs which are finite in the dimensional regulator $\epsilon$. We use the package {\tt Reduze} \cite{Panzer:2015ida,von_Manteuffel_2015,von_Manteuffel_2016,vonManteuffel:2012np} to identify these MIs. They can be evaluated efficiently for Euclidean kinematics using the method of SD, since only a small number of sectors is needed for finite integrals and no contour deformation is required to avoid Minkowskian thresholds. We employ the package {\tt pySecDec} \cite{Borowka:2017idc,Borowka:2018goh} for this purpose. The derivation of a DE system is done with the help of the IBP reduction program {\tt Kira} \cite{Maierhoefer:2017hyi,Klappert:2020nbg,Klappert:2019emp,Klappert:2020aqs}. With the boundary terms fixed numerically and the DE system derived analytically, we transport the Euclidean point to the Minkowski point with the aid of the method of series expansions of the DE system \cite{Pozzorini:2005ff,Moriello:2019yhu,Bonciani:2019jyb}\footnote{ See also Ref.~\cite{Fael:2021kyg} for a similar approach for single-scale problems.} as implemented in {\tt DiffExp} \cite{Hidding:2020ytt}.

\begin{figure}[h!]
\begin{center}
\includegraphics[scale=0.3]{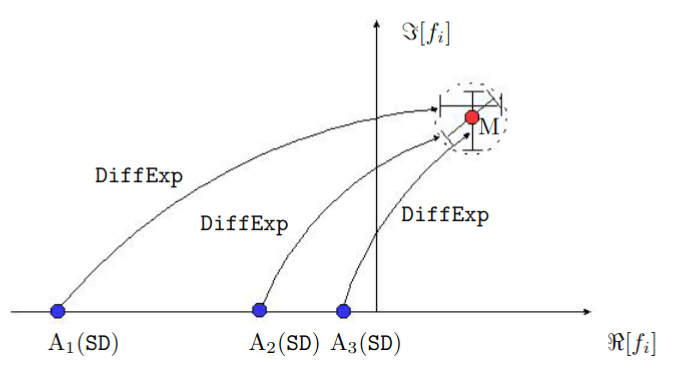}
\caption{Illustration of the DE transport method. The boundary conditions for the integral $f_i$ are evaluated at one or several Euclidean points $\rm A_k$, where the integral is purely real and one can obtain robustly converging numerical results with the SD method (using the package {\tt pySecDec} in our case). The boundary value(s) are transported to the physical kinematic point, using solutions of the DE system eq.~\eqref{eq:general-de} derived with {\tt DiffExp}, yielding the final result indicated by the red dot in the figure. The numerical uncertainty of a boundary value translates to an error estimate of the final result, as illustrated by the error bars in a zoomed-in area in the dotted circle, which permits a non-trivial cross-check if several boundary values $\rm A_k$ are employed. 
\label{fig:boundterms}
}
\end{center}
\end{figure}

As demonstrated in Fig.~\ref{fig:boundterms} we may choose different Euclidean points to fix the boundary terms numerically. This allows us to obtain a numerical error estimate of our automated method by taking the difference of two generated results for the same final Minkowski point. A more detailed discussion of the error estimate is provided in the supplemental material.

Typically, the transport from the Euclidean boundary point to the physical Minkowski kinematics requires several steps since the convergence radius of the series expansion at the boundary point is not large enough to reach the target point. The program {\tt DiffExp} automatically determines the convergence radius and the number of required transport steps.

In general, the complexity of the multi-loop computation increases with the number of loops and independent scales and the number of MIs involved. 
In our automated approach, the largest investment of computing resources is required for the IBP reduction with {\tt Kira} and the numerical evaluation of the boundary terms with {\tt pySecDec}. However, the former needs to be done only once for a given Feynman integral family, and the latter only once for a given choice of mass parameter values. The transport to the Minkowski region with {\tt DiffExp} is very fast, so that one can easily evaluate results for multiple different kinematic points, as needed e.g. for phase-space integrations. Quantitative information on the run time for our approach is given in the supplemental material. There the reader can also find a description for how our method can be extended to problems with multiple time-like momentum scales.

\paragraph{{\rm III.} Results and discussion.}  

To demonstrate the power and broad applicability of our method, in the following and in the supplemental material, we present examples for 3-loop self-energy and vertex integrals and 2-loop box integrals.
As discussed in the introduction, these are all examples of key theory ingredients for the physics program of future $e^+e^-$ colliders and/or the HL-LHC. The 3-loop integrals are needed for currently unknown third-order corrections to electroweak precision observables connected with Z-boson producton and decay, whereas two-loop box integrals are important to improve the precision of several $2\to 2$ processes, such as $W^+W^-$, $ZH$ or $f\bar{f}$ production \footnote{See Refs.~\cite{Song:2021vru,Liu:2021wks} for recent independent efforts on electroweak two-loop box integrals.}.

The technique described in this article allows one to compute the desired integrals to, in principle, arbitrary order in the dimension regularization parameter $\epsilon = (4-D)/2$ with multi-digit precision. To achieve a certain order $\epsilon^k$, some boundary terms need to be evaluated to higher orders $k'>k$ in $\epsilon$. The required order $k'$ is determined automatically from the IBP relations. For the examples shown below, some simple boundary-term integrals have to be computed to ${\cal O}(\epsilon^7)$, whereas no more than ${\cal O}(\epsilon^3)$ is needed for more complicated boundary terms. When evaluating the boundary terms with SD as implemented in {\tt pySecDec}, the computing time grows approximately linear with the order in $\epsilon$.

All of the following numerical examples are based on the input parameters given in the supplemental material.

\begin{figure}[h!]
    \centering
    \begin{tabular}{cl}
    \includegraphics[scale=0.23]{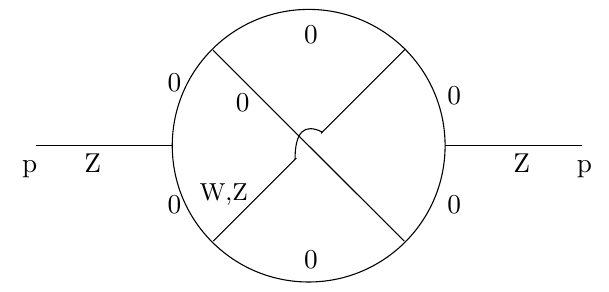} &
 \hspace*{-.5cm}   \includegraphics[scale=0.25]{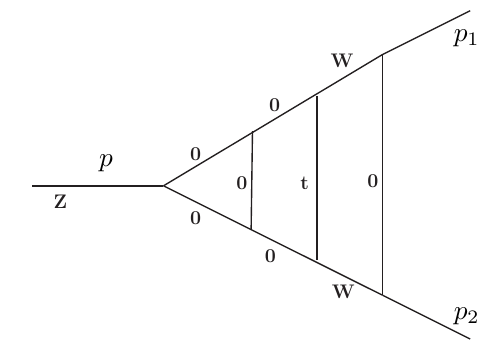} \\
  \hspace*{0cm}    lhNp1 &  \hspace*{1cm}  vtwPl \\
    \end{tabular}
    \caption{Three loop self-energy non-planar and planar vertex diagrams which correspond to integrals in \eqref{eq:lhnp} and \eqref{eq:vtwPl}, respectively. W, Z and t stand for the W-boson, Z-boson and top quark, respectively.}
    \label{fig:lhnp}
\end{figure}
{\underline{Example 1}}. As part of the 3-loop ${\cal O}(\alpha^2\as)$ corrections to electroweak precision observables, one encounters the following scalar non-planar self-energy integral with eight propagators and only one massive W- or Z-boson internal line \footnote{In this and the following examples, the large number of massless propagators occurs because all SM fermions except the top quark are taken to be massless.} (see Fig.~\ref{fig:lhnp} left):
\begin{align}
&I_{\text{lhNp1}}[D,\{a_{i}\},p^2,M_{a}^2]= \int\frac{\mathfrak{D}q_1\mathfrak{D}q_2\mathfrak{D}q_3}{[(q_1 - q_2)^2]^{a_1} [q_2^2]^{a_2}} \notag \\
&\times \frac{1}{[(q_1 - q_3)^2]^{a_3} [(q_2 - q_3)^2-M_a^2 ]^{a_4} [q_3^2]^{a_5}} \notag \\
&\times \frac{[q_1^2]^{-a_9}}{ [(q_1+p)^2]^{a_6} [(q_1-q_2+p)^2]^{a_7} [(q_3 + p)^2]^{a_8}}\,.
\label{eq:lhnp}
\end{align}
where $\mathfrak{D}q_n\equiv\frac{d^Dq_n}{i\pi^{D/2}}$ and $a=\rm W,Z$.
This example, for the parameter point $p^2=\MZ^2$ and $M_a=\MZ$ belongs to a group of integrals which are difficult to evaluate with SD due to threshold effects. Using {\tt pySecDec} with $10^7$ integration points we obtain a result with less than two digits precision:
\begin{align}
    &I_{\text{lhNp1}}^{{\tt pySecDec}}[4-2\epsilon,1,1,1,1,1,1,1,1,0,\MZ^2,\MZ^2]  \nonumber\\
    &= 0.460 - 19.164 \; i \pm ( 0.298 + 0.281 \; i ).
    \label{eq:lhnp-pySecDec}
\end{align}
Increasing the number of integration points does not improve the accuracy substantially. On the other hand, {\tt pySecDec} can deliver accurate results for Euclidean parameter points, $p^2 <0$, which are used as boundary terms for our automated DE transport. We thus obtain stable and precise results at the physical point:
\begin{align}
&I_{\text{lhNp1}}[4-2\epsilon,1,1,1,1,1,1,1,1,0,\MZ^2,\MZ^2]  \nonumber\\
&=-0.000000000
-19.1262302
\; i \nonumber\\
&\phantom{=} + (151.51529
- 150.40641
\; i)\; \epsilon + {\cal O}(\epsilon^2),
\label{eq:lhnp-MZ} \\[1ex]
&I_{\text{lhNp1}}[4-2\epsilon,1,1,1,1,1,1,1,1,0,\MZ^2,\MW^2] \nonumber\\ 
&= (5.1112260
- 18.5692007
\; i) \nonumber\\
&\phantom{=} + (194.660753
- 78.842016
\; i)\; \epsilon + {\cal O}(\epsilon^2).
\label{eq:lhnp-MW}
\end{align}
Here and in all the following results, we show all significant digits, i.e.\ the numerical error only affects digits beyond the ones shown in the equations. The error estimation will be described in more detail in the supplemental material.
The integral family $I_{\text{lhNp1}}$ \eqref{eq:lhnp} involves 30 master integrals and is considered simple in the context of our method.

\medskip
{\underline{Example 2}}. The next example is a family of 3-loop vertex integrals with one massive top quark and two massive W-boson propagators, see Fig.~\ref{fig:lhnp} (right), defined as
\begin{align}
&I_{\text{vtwPl}}[D,\{a_{i}\},p^2,M_\text{W}^2, m_\text{t}^2]= \int\frac{\mathfrak{D}q_1\mathfrak{D}q_2\mathfrak{D}q_3}{[q_3^2 - \MW^2]^{a_1} [q_2^2]^{a_2}} \notag \\ 
&\times \frac{1}{[q_1^2]^{a_3} [(q_1 - p)^2]^{a_4} [(q_2 - p)^2]^{a_5} [(q_3 - p)^2 - M_\text{W}^2]^{a_6}} \notag \\[.5ex]
&\times  \frac{[(q_1 - q_3)^2]^{-a_{10}}[(q_1 - p_2)^2]^{-a_{11}}[(q_2 - p_2)^2]^{-a_{12}}}{[(q_3 - p_1)^2]^{a_7} [(q_2 - q_3)^2 - m_\text{t}^2]^{a_8} [(q_1 - q_2)^2]^{a_9}}, 
\label{eq:vtwPl}
\end{align}
where $p=p_1+p_2$ and $p_1^2=p_2^2=0$. These integrals also appear in so far unknown  ${\cal O}(\alpha^2\as)$ corrections to Z-pole electroweak precision observables, constituting their most difficult parts.

With {\tt pySecDec} we are unable to obtain a numerical result for the Minkowski point $p^2=\MZ^2$. The problem already starts with the contour deformation which is necessary for SD with Minkowski kinematics and which fails to complete in a reasonable time. Similar to the SD method, the MB technique fails to deliver high-accuracy results for the considered integrals for $p^2=\MZ^2$. 

Using our automated DE transport method, the calculation  requires the numerical evaluation of 77 master integrals with Euclidean kinematics, $p^2<0$, for the boundary terms. For the purpose of the present example, they have been evaluated with {\tt pySecDec} to 10-digit accuracy. After the transport to the physical point $p^2=\MZ^2$, we get at least eight significant digits for integrals of the family \eqref{eq:vtwPl} up to tensor rank-3 (i.e. $-3\leq a_{10}+a_{11}+a_{12}\leq 0$). We here give numerical result for one rank-3 case:
\begin{align}
&I_{\text{vtwPl}}[1,1,1,1,1,1,1,1,1,-1,-1,-1,\MZ^2,\MW^2,\mt^2] = \nonumber\\
  &0.0833333333
  /\epsilon^3 
  +0.636273147
  /\epsilon^2 \nonumber\\ 
&+(0.63462699
  +0.77044487
  \;i)/\epsilon\nonumber\\
&+(5.5847828
  +6.1606031
  \;i) + {\cal O}(\epsilon),
\label{eq:vtwPl-1}
\end{align}
Additional examples, a 3-loop self-energy diagram with many massive propagators and a two-loop box diagram with four scales, are discussed in the supplemental material.

\paragraph{{\rm IV}. Summary and Outlook.}
 
In this work, we have proposed an efficient and versatile approach for the evaluation of a wide class of massive multi-loop, multi-scale Feynman integrals numerically, with typically 8 or more digits precision. It is
based on the method of DEs with boundary terms specified for Euclidean kinematics, which are transported to the physical Minkowski kinematics using series solutions of the DE. The Euclidean boundary term integrals avoid all threshold singularities and thus can be straightforwardly evaluated numerically. Our implementation combines the public programs {\tt Kira}, {\tt Reduze}, {\tt pySecDec} and {\tt DiffExp} in a way that allows us to automatically construct the required integral families and the transport from the Euclidean boundary point to the physical kinematic point. 

In principle, the technique can be extended to higher numerical accuracy and to wider classes of integrals with more loops and more external legs. A major bottleneck are the integration-by-parts (IBP) reductions that are needed to construct the DE system. A significant speed-up of this step is achieved when using numerical values for the relevant mass and kinematic parameters. In addition, the evaluation of the boundary terms for Euclidean kinematics can be time-consuming if a high level of precision is required. Fortunately, there are ongoing improvements to the sector decomposition (SD) and Mellin-Barnes (MB) methods, see e.g.~Refs.~\cite{Borowka:2018goh, Borinsky:2020rqs} and \cite{Ananthanarayan:2020fhl}.

It is worth mentioning that the 3-loop examples presented in this article are very difficult to solve with existing analytical techniques (e.g.~using IBP and DE) and general numerical methods (such as SD or MB methods). The proposed new technique is sufficiently general to provide the foundation for the computation of the required 3-loop corrections needed for electroweak and Higgs precision studies at the HL-LHC and future $e^+e^-$ colliders, which are key elements of the physics program of these machines \cite{Blondel:2018mad,Anastasiou:2016cez}. Other applications include flavor physics at Belle-II and low-energy precision tests of the Standard Model.
\paragraph{Acknowledgements}
This work has been supported in part by the Polish National Science Center (NCN) under grant 2017/25/B/ST2/01987, the Research Excellence Initiative of the University of Silesia in Katowice, the U.S.~National Science Foundation under grant nos.~PHY-1820760 and PHY-2112829, MH is supported by the European Research Council under ERC-STG-804286 UNISCAMP.

\section{Supplemental Material}
\subsection{Numerical results for relevant Master Integrals}

A set of auxiliary files are provided together with this article, which contain numerical results for the master integrals evaluated at the physical point (i.e.~for Minkowski kinematics). Here are the building blocks for computing the results in eqs.~(\ref{eq:lhnp-MZ}),(\ref{eq:lhnp-MW}),(\ref{eq:vtwPl-1}) in the main text and eqs.~(\ref{eq:taNpT1})-(\ref{eq:taNpT3}). The following files correspond to the different examples described in the text:

\medskip\noindent
\begin{tabular}{@{}ll@{}}
Example 1 (main text): \ &
\texttt{master\_lhNp1\_minkowski.m} \\ Example 2 (main text): \ &
\texttt{master\_vtwPl\_minkowski.m} \\ Example 3 (main text): \ &
\texttt{master\_box2l\_minkowski.m} \\ Example 4 (appendix): \ &
\texttt{master\_taNp1\_minkowski.m} \\ \end{tabular}

\medskip
The objects \texttt{lhNp1D4[...]}, \texttt{lhNp1D6[...]}, \texttt{lhNp1D8[...]}, \texttt{lhNp1D10[...]}, \texttt{taNp1D4[...]}, \texttt{taNp1D6[...]}, \texttt{vtwPlD4[...]}, \texttt{vtwPlD6[...]}, \texttt{vtwPlD8[...]} and \texttt{box2lD6[...]} define the master integrals for the four topologies \texttt{lhNp1}, \texttt{taNp1}, \texttt{vtwPl} and \texttt{box2l}, where \texttt{D4}, \texttt{D6}, \texttt{D8}, \texttt{D10} denote that the master integral is evaluated around $\epsilon=0$ for $d=\{4,6,8,10\}-2\epsilon$, respectively.

The notation which we use comes from chosen classification of the 3-loop self-energy (SE) and vertex SM diagrams, in general we have
\begin{verbatim}
ta = SE diagrams with top quark and photon
lh = SE diagrams with light quark(s) W/Z/H
vtw = 3-loop vertex topologies with top and W
Pl = planar ladder topologies
Np = non-planar topologies
\end{verbatim}

\subsection{Input parameters}

The input parameters for the \texttt{lhNp1}, \texttt{taNp1} and \texttt{vtwPl} topologies are summarized in Tab.~\ref{tab:input_parameters}, where each parameter $\MZ=\mz/\mz$, $\MW = \mw/\mz$ and $\mt = \lmt/\mz$ represents a physical scale rescaled by $\mz$. They are written as rational numbers suitable for IBP reductions with {\tt Kira}. The dimensionful values for $\mz$, $\mw$, $\lmt$  are the same as in Ref.~\cite{Dubovyk:2019szj}, i.e. $\mz = 91.1876$~GeV, $\mw = 80.385$~GeV, $\lmt = 173.2$~GeV.
\begin{table}[ht]
\renewcommand{\arraystretch}{1.6}
\caption{Input values, which are used for the 3-loop self-energy and vertex examples in this paper, are rescaled such that the Z-boson mass is 1.}
\label{tab:input_parameters}
\vspace{-2ex}
\begin{center}
\begin{tabular}{cc}
\hline
 Mass & Input value \\
\hline
 $\MZ$ & $1$   \\
 $\MW$ & $401925/455938 = 80.385/91.1876$ \\
 $\mt$ & $433000/227969 = 173.2/91.1876$ \\
\hline
\end{tabular}
\end{center}
\end{table}

\subsection{3-loop self-energy integrals with top quarks}

\begin{figure}[h!]
    \centering
    \includegraphics[scale=0.27]{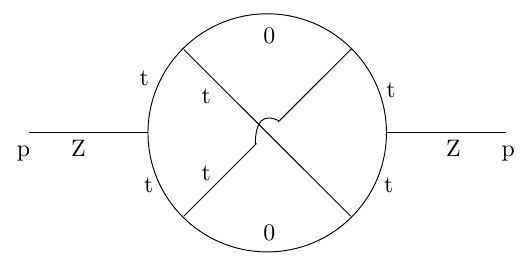}\\
    taNp1
    \caption{Three loop self-energy non-planar integral defined in (\ref{eq:taNp1}). $\text{Z}$  and $\text{t}$ stands for the massive SM Z gauge boson and the top quark, respectively.}
    \label{fig:lhnptops}
\end{figure}

\begin{align}
&I_{\text{taNp1}}[D,\{a_{i}\},p^2, m_\text{t}^2] = \int \mathfrak{D}q_1\mathfrak{D}q_2\mathfrak{D}q_3 \notag \\
&\frac{1}{[(q_1+p)^2]^{a_1} [(q_1 - q_2)^2 - \mt^2]^{a_2} [(q_1 - q_2 + p)^2 - \mt^2]^{a_3}} \notag \\
&\times \frac{1}{ [q_2^2 - \mt^2]^{a_4} [(q_1-q_3)^2 - \mt^2]^{a_5} [(q_2 - q_3)^2]^{a_6}} \notag \\
&\times \frac{[q_1^2]^{-a_{9}}[(q_2+p)^2]^{-a_{10}}}{[q_3^2 - \mt^2]^{a_7} [(q_3+p)^2 - \mt^2]^{a_8}}, 
\label{eq:taNp1}
\end{align}
see  Fig.~\ref{fig:lhnptops}, include the top quark which circulates through six propagators. These integrals are computed with at least ten digits accuracy in the Minkowski point  around $\epsilon=0$ for $D=4-2\epsilon$
\begin{align}
&I_{\text{taNp1}}[1,1,1,1,1,1,1,1,-2,-1,\MZ^2,\mt^2]= \notag \\
&8.27490485938
/\epsilon^3 - 34.9869281045
/\epsilon^2 \notag \\
&+102.43077689
/\epsilon - 253.5072352
,
\label{eq:taNpT1} \displaybreak[0] \\[1ex]
&I_{\text{taNp1}}[4-2\epsilon,1,1,1,1,1,1,1,1,-1,-2,\MZ^2,\mt^2]= \notag \\
&9.47745432492
/\epsilon^3 - 40.4955852564
/\epsilon^2 \\
&+ 116.63419570
/\epsilon - 273.3763275
,
\label{eq:taNpT2} \displaybreak[0] \\[1ex]
&I_{\text{taNp1}}[1,1,1,1,1,1,1,1,0,-3,\MZ^2, \mt^2]= \notag \\
& 19.8715753165
/\epsilon^3 - 74.436608700
/\epsilon^2 \\
&+ 239.02713087
/\epsilon - 540.2221570
.
\label{eq:taNpT3}
\end{align}
Note the tenth propagator in (\ref{eq:taNp1}) is linearly dependent and can be written in terms of the first nine propgators. We included it as an auxiliary propagator to the definition to improve the readability of the final results in (\ref{eq:taNpT1})-(\ref{eq:taNpT3}).

\subsection{Two-loop box diagram}
\begin{figure}[h!]
    \centering
    \includegraphics[scale=0.26]{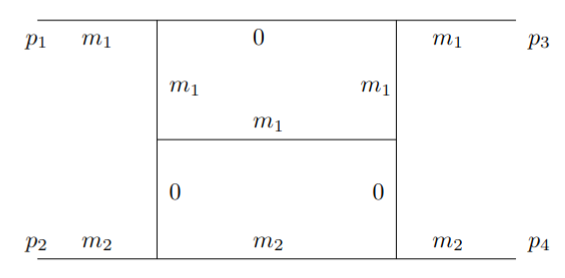}
    \caption{Two-loop box diagram with four scales: $s,t,m_1,m_2$.}
    \label{fig:2lboxm1m2}
\end{figure}
We consider a family of box integrals that are part of the ${\cal O}(\alpha^2)$ corrections to massive  $e\mu$ scattering \cite{Banerjee:2020tdt}, see Fig.~\ref{fig:2lboxm1m2}:
\begin{align}
&I_{\text{box2l}}[D,\{a_{i}\},s,t, m_{1}^{2},m_{2}^{2}]= \int \frac{\mathfrak{D}q_1\mathfrak{D}q_2}{[(q_1-p_1)^2]^{a_1} [q_1^2 - m_1^2]^{a_2}} \notag \\
&\times \frac{1}{[(q_1 - p_1 + p_3)^2 - m_1^2]^{a_3} [(q_1 - q_2)^2 - m_1^2]^{a_4} } \notag \\
&\times\frac{[(q_{2}-p_{1})^2]^{-a_{8}}[(q_{1}-p_{2})^2]^{-a_{9}}}
{[(q_2 - p_1 + p_3)^2]^{a_5} [(q_2 + p_2)^2-m_2^2]^{a_6} [q_2^2]^{a_7}}
\label{eq:box2l}
\end{align}
where $p_1^2=p_3^2=m_1^2$, $p_2^2 = p_4^2 = m_2^2$, $p_1\,p_2=(s-m_1^2 - m_2^2)/2$, $p_1\,p_3 = (2 m_1^2 - t)/2$, $p_2\,p_3 = (s + t - m_1^2 - m_2^2)/2$. With input parameters $s=2$, $t=5$, $m_1^2=4$ and $m_2^2=16$ we obtain
\begin{align}
&I_{\text{box2l}}[2,1,1,1,1,1,1,0,0,s,t,m_1^2,m_2^2]= \notag \\
&+ 0.000328707579
/\epsilon^2 \notag \\
&- (0.0014129475
- 0.0020653306
\;i)/\epsilon \notag \\
&- (0.005702737
- 0.000485980
\;i) + {\cal O}(\epsilon).
\label{eq:box2l-1}
\end{align}
The $I_{\text{box2l}}$ integral family involves 55 master integrals. The input parameters for the \texttt{box2l} topology are $s=2$, $t=5$, $m_1^2=4$ and $m_2^2=16$.

\subsection{Tracing the run time of the calculations}
For the calculations demonstrated in this paper it makes sense to categorize the run time in three parts. The first part includes the preparation of the DE system suitable for later use with {\tt DiffExp}. For this step, the major bottleneck is the run time of {\tt Kira}. For the most complicated example {\tt vtwPl} we need 4 hours to construct the DE system on a 12 core, 2.7\,GHz Intel Xenon processor cluster with 128\,GB of RAM. The second part is the computation of Euclidean boundary terms, which we evaluate with the sector decomposition program {\tt pySecDec}. Since the boundary terms are finite integrals by construction and are free of thresholds in the Euclidean region, the Monte-Carlo (MC) or quasi-Monte-Carlo (QMC) scaling rules apply straightforwardly. For QMC the theoretical bound on the error is proportional to $\mathcal{O}(1/N)$, where $N$ is the number of points. For traditional MC, the bound is $\mathcal{O}(1/\sqrt{N})$. For the most complicated example {\tt vtwPl}, we ran the QMC integrator configured with a maximum of $10^7$ points. We ran the computation on a machine equipped with a 16 core Threadripper Pro 3955WX. Two different points in the Euclidean region were chosen, and both took approximately 3 days to complete. The maximum relative error reported by pySecDec was approximately $2 \cdot 10^{-9}$. We expect that the precision of our result can be further improved by running on a more powerful machine, or by allocating a longer running time. The final part of the computation is the transport of the boundary terms to the Minkowski regions. The run time of this part of the computation depends on the size of the differential equations, the order of the homogeneous parts of the differential equations, and the number of line segments. In our case we are interested in the evaluation of the Feynman integral in one particular point. With {\tt DiffExp} this is accomplished in about 3.5 hours on a single CPU core for the most complicated example {\tt vtwPl}. Note that once the computation with {\tt DiffExp} is finished the integral is available for a fast evaluation (in terms of milliseconds) at an arbitrary point along the line of the transport from Euclidean to Minkowski point.

\subsection{Tracing the error of the calculations}
    The numerical error estimation has two main ingredients, the numerical series expansion of the differential equation (DE) system with {\tt DiffExp} and the numerical evaluation of the boundary terms with {\tt pySecDec}. The numerical error from the series expansion can always be rendered negligible compared to the error from the boundary terms by evaluating the expansion to sufficiently high order. On the other hand, the numerical errors of the initial boundary terms can usually be directly mapped to the final result. However, instead of relying on the error estimate from {\tt pySecDec}, we verify the accuracy by carrying out separate transports from two different initial boundary points to the same final Minkowski point and taking the difference as an error estimate. This cross-check was sufficient for the examples presented here. Also, if necessary,  it is relatively straightforward to increase the accuracy of the boundary terms by using more Monte-Carlo integration points, since the SD integrand is well-behaved in the Euclidean region.

\subsection{Iterative approach for multi-scale kinematics}

In the case of problems with multiple time-like momentum scales, one may use our method in an iterative way. For instance, we may derive the DE system just in one scale $m_1$, while setting all other scales $\{m_i\} \setminus  m_1$ to numerical values, to simplify the IBP reduction. Then one can carry out the transport with respect to $m_1$ from the numerical point $a_1$, where the boundary conditions are supplied with all scales taken to be Euclidean (space-like), to a new point of interest denoted as $a_2$, where $m_1$ is time-like. Next we derive a new set of DEs with respect to the next scale $m_2$, and again we set all other scales $\{m_i\} \setminus m_2$ to numerical values, and transport from the point $a_2$ to a new point of interest $a_3$.
One can continue in this way until all ($n$) relevant momentum scales have been transported from the Euclidean to the time-like domain. The downside of the iterative approach is that one has to repeat the IBP reduction whenever we want to reach a new kinematic point that differs in any of the $n-1$ first transported variables. Thus new improvements in the IBP reduction programs are highly anticipated.


\end{document}